\documentclass[twocolumn, usenatbib]{mnras}
\usepackage{newtxtext,newtxmath}

\usepackage[T1]{fontenc}
\usepackage{xcolor}
\usepackage{lineno}
\usepackage[normalem]{ulem}

\DeclareRobustCommand{\VAN}[3]{#2}
\let\VANthebibliography\thebibliography
\def\thebibliography{\DeclareRobustCommand{\VAN}[3]{##3}\VANthebibliography}

\usepackage{graphicx}	
\usepackage{amsmath}	

\newcommand\Suzaku{{\textit Suzaku }}

\title[X-ray Flux and Spectral Variability of OJ 287]{X-ray Flux and Spectral Variability of the Blazar OJ 287 with \it Suzaku}

\author[Dongtao Zhou et al.]{Dongtao Zhou,$^{1,2}$
    Zhongli Zhang,$^{1,3}$\thanks{E-mail: zzl@shao.ac.cn}
    Alok C.\ Gupta,$^{4,5,6}$
    Pankaj Kushwaha,$^{7}$
    Paul J.\ Wiita,$^{8}$
    Minfeng Gu,$^{6}$
    \newauthor
    and Haiguang Xu$^{9}$
\\
\\
$^{1}$Shanghai Astronomical Observatory, Chinese Academy of Sciences, Shanghai 200030, China\\
$^{2}$School of Astronomy and Space Science, University of the Chinese Academy of Sciences, Beijing 100012, China\\
$^{3}$Key Laboratory of Radio Astronomy and Technology, Chinese Academy of Sciences, A20 Datun Road, Chaoyang District, Beijing 100101, China\\
$^{4}$Xinjiang Astronomical Observatory, CAS, 150 Science-1 Street, Urumqi 830011, People's Republic of China \\
$^{5}$Aryabhatta Research Institute of Observational Sciences (ARIES), Manora Peak, Nainital 263001, India\\
$^{6}$Key Laboratory for Research in Galaxies and Cosmology, Shanghai Astronomical Observatory, Chinese Academy of Sciences, 80 Nandan Road, \\
~~Shanghai 200030, China\\
$^{7}$Indian Institute of Science Education and Research (IISER) Mohali, Knowledge city, Sector 81, SAS Nagar, Manauli 140306, India\\
$^{8}$Department of Physics, The College of New Jersey, 2000 Pennington Rd., Ewing, NJ 08628-0718, USA\\
$^{9}$School of Physics and Astronomy, Shanghai Jiao Tong University, Shanghai 200240, China
}

\date{Accepted 2024 July 9. Received 2024 July 4; in original form 2024 May 8}

\pubyear{2024}

\begin{document}
\label{firstpage}
\pagerange{\pageref{firstpage}--\pageref{lastpage}}
\maketitle

\begin{abstract}
We present analyses of \Suzaku XIS light curves and spectra of the BL Lac object OJ 287 with observations positioned primarily around proposed recurrent optical outbursts. The first two observations were performed in 2007 April 10 -- 13 (epoch 1) and 2007 November 7 -- 9 (epoch 2) that respectively correspond to a low and a high optical state and which, within the binary supermassive black hole model for OJ 287, precede and follow the impact flare. The last three observations, made consecutively during 2015 May 3 -- 9 (epoch 3), were  during the post-impact state of the 2013 disc impact and are the longest continuous X-ray observation of OJ 287 taken before the optical outburst in 2015 December. Intraday variability is found in both the soft (0.5 -- 2 keV) and hard (2 -- 10 keV) bands. The discrete correction function analysis of the light curves in both bands peaks at zero lag during epochs 2 and 3,  indicating that the emission in both bands was cospatial and emitted from the same population of leptons. Power spectral densities of all three light curves are red noise dominated, with a rather wide range of power spectrum slopes. These X-ray spectra are overall consistent with power-laws but with significantly different spectral indices. In the 2015 observations the X-ray spectrum softens during the flare, showing an obvious soft X-ray excess that was not evident in the 2007 observations. We discuss the implications of these observations on the jet, the possible accretion disc, and the binary supermassive black hole model proposed for the nearly periodic optical flaring of OJ 287.
\end{abstract}




\section{Introduction} 
\label{sec:introduction}
The blazar subclass of radio-loud (RL) active galactic nuclei (AGNs) is the common name for the union of BL Lacertae objects (BLLs) and flat spectrum radio quasars (FSRQs). Blazars host a  central supermassive black hole (SMBH) in the mass range  $\sim \rm{10}^{6} - \rm{10}^{10} \rm{M}_{\odot}$  \citep[e.g.,][]{1984ARA&A..22..471R} and emit Doppler-boosted radiation from a large-scale relativistic plasma jet pointing nearly towards the line of sight of the observer \citep[e.g.,][]{1995PASP..107..803U}. Blazars emit radiation across the whole electromagnetic (EM) range from radio to $\gamma$-ray bands, display strong flux, polarization and spectral variability on diverse timescales that range from a few minutes to several years or even decades. Broadly, blazar variability has been classified into three subclasses: intraday variability (IDV) in which the flux noticeably varies over timescales of a few minutes to less than a day  \citep{1995ARA&A..33..163W}, short-term variability (STV) with variability timescales from days to weeks, and long-term variability (LTV) with timescales of months or longer \citep{2004A&A...422..505G}.\\
\\
Blazar broadband continuum emission is almost entirely non-thermal and exhibits a characteristic bimodal spectral energy distribution \citep[SED; e.g.,][]{1998MNRAS.299..433F}. The low energy hump of the SED, extending from radio up to  X-ray energies, is dominated by synchrotron radiation from relativistic leptons, whereas the high energy hump peaking at MeV--GeV $\gamma$-rays and sometimes extending down to X-rays, can arise by either inverse Compton (IC) or hadronic processes \citep[e.g.,][and references therein]{1983ApJ...264..296M,1998A&A...333..452K,2003APh....18..593M,2004NewAR..48..367K,2007Ap&SS.309...95B,2013ApJ...768...54B,2017SSRv..207....5R}. Based on the location of the first peak (synchrotron peak) of their SEDs, $\nu_{\rm peak}^{\rm syn}$, blazars have historically been divided into low-energy-peaked blazars (LBLs) and high-energy-peaked blazars (HBLs) \citep{1995ApJ...444..567P}.  In LBLs, $\nu_{\rm peak}^{\rm syn}$ lies in the near-infrared (NIR)/optical bands, and for HBLs $\nu_{\rm peak}^{\rm syn}$ fall in the ultraviolet (UV) or X-rays bands. The second component generally peaks at GeV and TeV energies for LBLs and HBLs, respectively \citep{1995ApJ...444..567P,1995A&AS..109..267G}. \citet{2010ApJ...716...30A} extended the blazar classification based on their $\nu_{\rm peak}^{\rm syn}$ to: LSPs (low-synchrotron-peaked blazars) with $\nu_{\rm peak}^{\rm syn} \leq \rm{10}^{14}$ Hz; ISPs (intermediate-synchrotron-peaked blazars) with $\rm{10}^{14} \rm{Hz} < \nu_{\rm peak}^{\rm syn} < \rm{10}^{15}$ Hz; and HSPs (high-synchrotron-peaked blazars) with $\nu_{\rm peak}^{\rm syn} \geq \rm{10}^{15}$ Hz.\\ 
\\
OJ 287\footnote{\url{https://www.lsw.uni-heidelberg.de/projects/extragalactic/charts/0851+203.html}} at $z=0.3056$ \citep{1985PASP...97.1158S} is a BL Lac type of blazar.
In terms of broadband SED, it is a LBL/LSP source with the synchrotron component peaking around NIR energies \citep[e.g.][]{2010ApJ...716...30A,2020Galax...8...15K,2022JApA...43...79K}. OJ 287 has been observed in optical bands since 1888 and using this century-long, but rather sparse light curve (LC), \citet{1988ApJ...325..628S} argued for the first time that the blazar displayed 
an optical quasi-periodic oscillation (QPO) pattern with a period of $\sim$ 12 yr. 
Hence, an extensive international observing campaign for optical bands called OJ-94 was organized and the predicted outbursts were really observed, with a double peaked structure, in which the separation between the two peaks was $\sim$ 1.2 yr \citep{1996A&A...305L..17S,1996A&A...315L..13S}. 
The presence of the $\sim$ 12 yr structure led \citet{1996ApJ...460..207L} to propose a double-outburst structure involving the impacts of a secondary supermassive black hole (SMBH) on the accretion disc of the primary one, and this was immediately verified by \citep{1996A&A...315L..13S}.
In the subsequent observing campaigns of OJ 287 carried out during 2005–-2007, and 2015--2019, the pair of outbursts were detected at the ends of 2005 and 2007, i.e., separated by $\sim$2 yr \citep{2009ApJ...698..781V}, and in 2015 December and 2019 July, i.e. separated by $\sim$ 3.5 yr, respectively, \citep{,2009PASJ...61.1011S,2016ApJ...819L..37V,2017MNRAS.465.4423G,2020ApJ...894L...1L}. This 12-yr periodicity of the main flare structure has persisted even as the historical database has increased roughly a hundred-fold as new data have been collected over several decades \citep[e.g.,][in which the optical data have been given in digital form for the first time]{2024ApJ...968L..17V}. \\ 
\\
The brightness of this source and especially the presence of these  $\sim$ 12-yr recurring optical outbursts  
has triggered extensive and intensive investigations of OJ 287 \citep[e.g.,][and references therein]{1996A&A...305L..17S,1996A&A...315L..13S,2016ApJ...819L..37V,2020MNRAS.498L..35K,2020Galax...8...15K,2022JApA...43...79K}. 
It shows strong and frequent flux variations,  with the majority of them accompanied  by significant spectral changes, especially at X-ray energies \citep[e.g.,][and references therein]{2022MNRAS.509.2696S,2022JApA...43...79K}. \citet{2009PASJ...61.1011S} performed spectral analysis of two of the {\it Suzaku} observations of OJ 287 taken in 2007, which we have also analyzed in the present study. They found simple power laws fit the spectra well. Existing data and broadband spectral studies during diverse flux states indicate that blazars rarely show significant spectral evolution \citep[e.g.,][and references therein]{2004A&A...413..489M,2008A&A...478..395M,2018MNRAS.473.3638G,2020Galax...8...62G},  but OJ 287 was recently reported to be in a new activity state that lasted for about 4 years \citep{2018MNRAS.473.1145K, 2018MNRAS.479.1672K, 2020MNRAS.498L..35K}, that was in coincidence with the claimed $\sim 12$-yr recurring optical outburst \citep[e.g.,][and references therein]{2020Galax...8...15K,2022JApA...43...79K}. Its X-ray spectrum has been one of the most diverse, spanning all the behaviours reported for the entire blazar class \citep[e.g.][]{2020Galax...8...15K,2022JApA...43...79K,2022MNRAS.509.2696S}, including extremely soft \citep{2020MNRAS.498L..35K,2018MNRAS.479.1672K,2021ApJ...921...18K,2022MNRAS.509.2696S}, and intermediate \citep{2018MNRAS.479.1672K,2017MNRAS.468..426S,2022MNRAS.509.2696S}, as well as its apparently usual, low-flux, spectrum \citep{2010ApJ...716...30A,2013MNRAS.433.2380K,2018MNRAS.473.1145K,2022MNRAS.509.2696S}. In addition, it has shown excess emission in X-rays that is closer to the behaviours seen in radio-quiet AGNs, normally referred to as soft X-ray excess \citep{2001PASJ...53...79I,2020ApJ...890...47P}. \\
\\
Among the most puzzling variations in AGNs are those happening on the IDV timescales. Rapid IDV was first discovered in the X-ray band in 1978 \citep{1978ApJ...225L.115M} and later, with the improvement of X-ray detectors, was found to be a common property of radio-quiet AGNs \citep[e.g.,][and references therein]{1987Natur.325..694L,1987Natur.325..696M,2002MNRAS.332..231U,2004MNRAS.348..783M,2009A&A...494..905P,2010A&A...510A..65P,2019MNRAS.485.1454P}. For these radio-quiet AGNs the level of variability was found to strongly depend on the SMBH mass \citep{2004ApJ...617..939M,2004MNRAS.348..207P,2021Sci...373..789B}. Blazars, on the other hand, have been found to exhibit flux/brightness variations on all time scales probed, from sub-minutes to decades \citep[e.g.,][and references therein]{2015MNRAS.451.1356K,2017ApJ...841..123P,2018MNRAS.480.4873A,2022ApJ...939...80D}. The variations are primarily stochastic over long-terms, and while they share statistical behaviors broadly similar to those of accretion-powered sources \citep{2015SciA....1E0686S}, they lack any clear relation to the inferred central engine properties \citep{2016ApJ...822L..13K,2017ApJ...849..138K}. The short-term statistical behaviors, on the other hand, seem to be different and complex \citep{2013ApJ...766...16E,2020Galax...8...66K}. \\
\\
The current work is part of an extensive project started in 2009 focusing on a detailed and systematic study of blazars’ X-ray flux and spectral variabilities that included searches for QPOs on IDV timescales. For these investigations we have extensively used public archive data of blazars taken from various X-ray satellites, e.g., {\it XMM-Newton, NuStar, Chandra}, and {\it Suzaku}. The outcomes and inferences from these studies using the {\it XMM-Newton} facility data have been reported in \citet{2009A&A...506L..17L,2010ApJ...718..279G,2014MNRAS.444.3647B,2016NewA...44...21B,2015MNRAS.451.1356K,2016MNRAS.462.1508G}, and \citet{2021MNRAS.506.1198D}. {\it Chandra} based results are reported in \citet{2018MNRAS.480.4873A} while {\it Suzaku} data have been published in \citet{2019ApJ...884..125Z,2021ApJ...909..103Z}, and the {\it NuStar} based findings of \citet{2017ApJ...841..123P,2018ApJ...859...49P} are summarized by us in \citet{2022MNRAS.511.3101P}. We also have carried out additional studies of blazars based on {\it XMM-Newton} observations, reporting results on spectral variability \citep{2017ApJ...850..209G,2018MNRAS.473.3638G,2017MNRAS.469.3824K} and temporal variability \citep{2022MNRAS.511.3101P,2022ApJS..262....4N,2022ApJ...939...80D}. \\
\\
To understand further the puzzling issue of blazar IDV, here we report on the five pointed {\it Suzaku} X-ray observations of the famous binary SMBH blazar candidate OJ 287 \citep[e.g.][and references therein]{1988ApJ...325..628S,2008Natur.452..851V,2012MNRAS.427...77V,2016ApJ...819L..37V,2024ApJ...968L..17V}. OJ 287 is one of the potential close binary SMBH candidates for the direct detection of gravitational wave (GW) emission by the Pulsar Timing Array (PTA) or an interferometer in space \cite[e.g.,][]{2018MNRAS.481.2249C,2019BAAS...51c.123B,2021ExA....51.1441B,2019A&ARv..27....5B}. These observations were taken during April and December 2007 and May 2015. In each of the pointed observations, the elapsed time ranges from $\sim$ 208 to 250 ks. We searched for IDV in flux and spectra, and also carried out power spectrum density (PSD) analysis to see whether or not they change and if there is any quasi-periodicity in the data. \\
\\
The paper is structured as follows: in Section 2, we discuss the {\it Suzaku} archival data and their reduction. Our various data analysis techniques are explained in Section 3. In Section 4 we provide the results. In Sections 5 and 6, a discussion and the conclusions of the paper are respectively presented.

\section{{\it Suzaku} Archival Data Reduction} 
\label{sec:Obeservation}
\begin{figure*}
    \center
	\includegraphics[width=\linewidth]{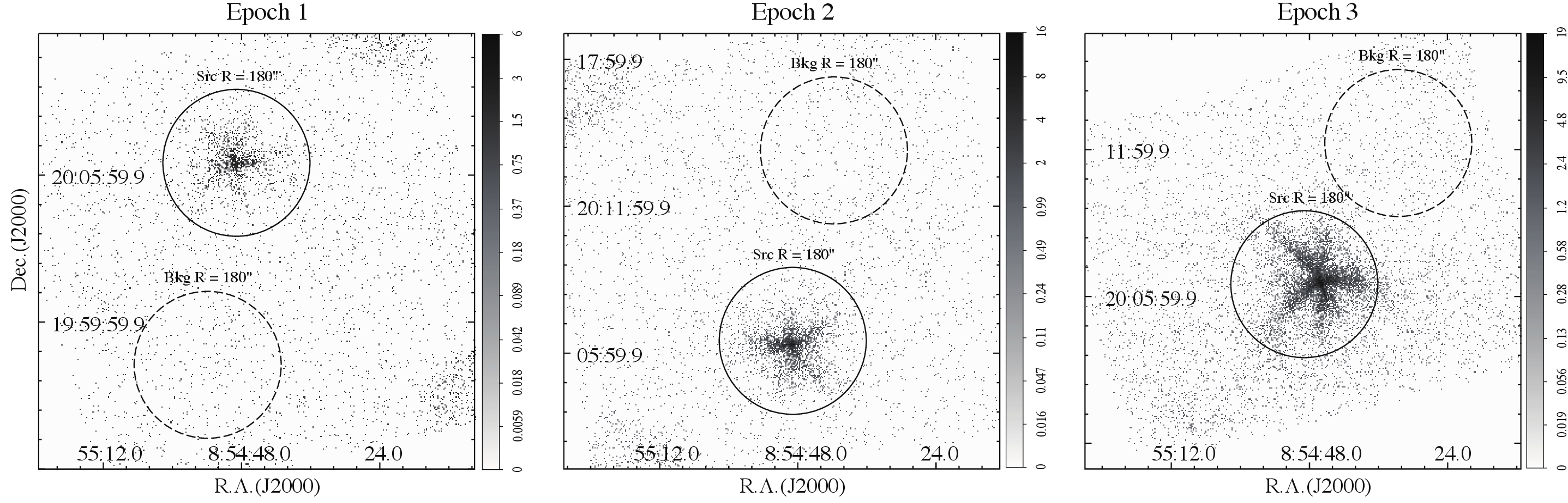}
    \caption{{\it Suzaku} observation images of the OJ 287 field made with XIS 0 without window option. The figures are on logarithmic scales with inverted gray colors. The source region in each epoch is a circle with radius of $180^{\prime\prime}$ centred at R.A. = $08^{h}54^{m}48.^{s}9$ and Dec. = $20^{\circ}06^{\prime}31^{\prime\prime}$. The background regions are chosen to away from the source and close to the edges of the CCD and they are measured within the same radius.}
    \label{fig:ccd}
\end{figure*}

{\it Suzaku} is a Japanese X-ray observatory in a low earth orbit with an apogee of 568 km and an orbital period of 5752 s.
The occlusion by the earth and the interruption of data acquisition when passing through the South Atlantic Anomaly (SAA), mean that the observing efficiency of {\it Suzaku} is normally less than $50\%$. But {\it Suzaku} is still an extremely good instrument for studying astronomical objects with high energy emission because of its excellent X-ray sensitivity throughout its broadband energy range of 0.2 -- 600.0 keV \citep{2007PASJ...59S...1M}. \\
\\
The X-ray Imaging Spectrometer \citep[XIS;][]{2007PASJ...59S..23K} was carried by {\it Suzaku} for measurements in its low-energy band (0.2--12 keV). The XIS contains four CCDs (charge-coupled devices) named XIS 0 to 3. XIS 0, 2, and 3 are front-illuminated (FI) CCDs and XIS 1 is back-illuminated (BI). The former were better calibrated, but are less sensitive than the latter in the soft X-rays below $\sim$1 keV. Pointed observations by {\it Suzaku} XIS on OJ 287 were carried out on five occasions in three epochs (Table \ref{tab:observation}): epoch 1 on  10 April 2007 (ID 702009010); epoch 2 on 7 November  2007 (ID 702008010); and epoch 3 on 3 May 2015 (ID 710011010\ref{fig:curve}), 6 May 2015 (ID 710011020), and 9 May 2015 (ID 710011030). Thus during epoch 3 \Suzaku was continuously observing OJ 287 for $\sim$ 750 ks ($\sim$ 9 days), making it one of the longest and most evenly sampled observations, and providing some of the best time resolution in any emission band for OJ 287 until now. \\
\\
We utilized and combined measurements from XIS 0 and 3 for the source IDV studies because XIS 2 stopped working after 9 November 2006. Data from XIS 1 were added only for the unified spectral fitting performed in section \ref{sec:spectra}. Data were processed with HEAsoft{\footnote{\url{https://heasarc.gsfc.nasa.gov/docs/software/heasoft/}}} \citep[v6.29c;][]{1995ASPC...77..367B}. The source signals were extracted in a total energy range of 0.5 -- 10.0 keV during the good time intervals (GTIs) of the five observations, taken within the solid circles with radii of $180^{\prime\prime}$, as shown in Figure \ref{fig:ccd}. Background regions were chosen to lie in the circles of the same radii near the far edges of the CCDs, and these count rates are $\sim15.4\%$, $\sim8.1\%$ and $\sim6.6\%$ of the source rates, respectively in each observation (Table \ref{tab:observation}). So these backgrounds are not negligible, especially for epoch 1 when the source is the dimmest, and we subtracted them from the source rates in further analyses. \\ 
\\
Setting the time binning to be 5752s, which is also the orbital period of {\it Suzaku}, we plotted the source background-subtracted light curves of the three epochs in: the total energy  band of 0.5--10.0 keV; the soft band of 0.5--2.0 keV; and the hard band of 2.0--10.0 keV in Figure \ref{fig:curve}. To characterize the spectral variations of X-ray emission, the hardness ratio (HR) is  often used \citep[e.g.][and references therein]{2017ApJ...841..123P,2018MNRAS.480.4873A,2019ApJ...884..125Z,2021ApJ...909..103Z}. The HR and its error $\sigma_{\rm HR}$ were defined as explained in \citet{2019ApJ...884..125Z}, and are plotted in the bottom panels of  Figure \ref{fig:curve}. Our following analyses are mainly based on these data.

\begin{table*}
	\centering
	\caption{The {\it Suzaku} observations of OJ 287}
	\label{tab:observation}
	\begin{tabular}{cccccccc} 
		\hline
        epoch    & ObsID     & Date       & MJD     & Elapse$^{a}$ & GTI$^{b}$ & Src Rate$^{c}$  & Bkg Rate$^{d}$     \\ 
                 &           &            &         & (ks)         & (ks)     & (count s$^{-1}$) & (count s$^{-1}$)  \\
        \hline
        1        & 702009010 & 2007-04-10 &  54200  &  225.3       &  102.8   &    0.188         &  0.029            \\
        2        & 702008010 & 2007-11-07 &  54411  &  208.5       &  112.1   &    0.358         &  0.029            \\
        3        & 710011010 & 2015-05-03 &  57145  &  247.0       &  111.6   &    0.384         &  0.024            \\
                 & 710011020 & 2015-05-06 &  57148  &  254.6       &  110.3   &    0.428         &  0.030            \\
                 & 710011030 & 2015-05-09 &  57151  &  249.0       &  107.7   &    0.377         &  0.025            \\
		\hline
        \noindent
        \textbf{Notes.}\\
        \multicolumn{4}{l}{$^a$ Total elapsed time of the observation.}\\
        \multicolumn{4}{l}{$^b$ Total clean Good Time Interval of the observation.}\\
        \multicolumn{4}{l}{$^c$ Source count rate in 0.5 -- 10 keV energy range of CCDs XIS 0+XIS 3.}\\
        \multicolumn{4}{l}{$^d$ Background count rate in 0.5 -- 10 keV energy range of CCDs XIS 0+XIS 3.}\\
	\end{tabular}

\end{table*}

\section{Analysis Techniques}
\label{sec:analysis}
To quantify the X-ray flux and spectral variability of the blazar OJ 287, we have used various standard data analysis techniques which we now briefly discuss.

\subsection{Excess Variance}

Blazars are generally characterized by rapid X-ray variability. In a blazar light curve, there will be some inborn experimental noise which produce finite errors, $\sigma_{err, i}$, for each of the  measurements that contribute additional variance to the observed variance \citep[e.g.][]{2017ApJ...841..123P,2018MNRAS.480.4873A}. The quantitative measurements of strength of true variance (intrinsic flux variability) are known as excess variance, $\sigma_{XS}$, and the fractional $rms$ variability amplitude, $F_{var}$ \citep{2002ApJ...568..610E}. We followed \citet{2003MNRAS.345.1271V} to define and compute $F_{var}$ and its error, as we did in our earlier work on the TeV emitting blazars Mrk 421 and PKS 2155$-$304 \citep{2019ApJ...884..125Z,2021ApJ...909..103Z}.

\subsection{Flux Variability Timescale}

We briefly discuss the IDV timescale estimation method which is explained in more detail elsewhere \citep[e.g.][]{2018A&A...619A..93B,2019ApJ...884..125Z}. \citet{1974ApJ...193...43B} introduced expressions for a weighted variability timescale $\tau_{var}$, and its errors $\Delta \tau_{var}$. We use the standard error propagation method \citep[similar to Equation 3.14 given in][]{2003drea.book.....B} to estimate the error: 

\begin{equation}
    \tau_{var}=  \left | \frac{\Delta t}{ \Delta ln(F_1/F_2)} \right |~;
\end{equation}

\begin{equation}
\Delta \tau _{var}
\simeq \sqrt{\frac{F_{1}^{2} \Delta F_{2}^{2} +F_{2}^{2} \Delta F_{1}^{2}}{F_{1}^{2}F_{2}^{2}\left (  ln \left [ F_{1}/F_{2} \right ]\right )^{4}}}\ \Delta t ~.
\end{equation}
Here $\Delta t$ refers to time interval between the measurements of variable flux $F$ \citep[see also][]{2008ApJ...672...40H},  $F_{1}$ and $F_{2}$ are the X-ray fluxes (in count sec$^{-1}$) which are used to estimate the shortest variability timescales, and $\Delta F_{1}$ and $ \Delta F_{2}$ are their corresponding errors. 

\subsection{Discrete Correlation Function}

The classical correlation function (CCF) requires evenly sampled time series LC data. Of course, most astronomical data are unevenly sampled, so \citet{1988ApJ...333..646E} introduced the discrete correlation function (DCF) which can work with such data. \citet{1992ApJ...386..473H} generalized the DCF method to include a better error estimate. We have provided a detailed description of the implementation of the DCF for {\it Suzaku} data in \citet{2019ApJ...884..125Z}. In general we have found that most of the DCFs between soft and hard X-ray bands in blazars are rather broad \citep[e.g.,][and references therein]{2019ApJ...884..125Z,2021ApJ...909..103Z,2022ApJS..262....4N}, 
so we fit them with a Gaussian function:
\begin{equation}
\label{equ:10}
DCF(\tau)=a \times {\rm exp}\Bigl[\frac{-(\tau - m)^{2}}{2 \sigma^{2}}\Bigr],
\end{equation}
where $a$, $m$, and $\sigma$ are the peak value of the DCF, the time lag at which the DCF peaks, and the width of the Gaussian function, respectively. 

\subsection{Power Spectral Density}

The power spectral density (PSD) is a periodogram analysis tool that finds the distribution of variability power as a function of temporal frequency and can be used in searching for possible quasi-periodic oscillations (QPOs). PSD analysis involves calculating the Fourier transform of the LC and then usually fitting it to a power-law, as AGN  PSDs are usually dominated by red-noise. A possible QPO may be claimed in a LC if the significance of any peak of PSD rising above the red-noise is $3\sigma$ (99.73\%) or more. We follow the approach given by \citet{2005A&A...431..391V} so that PSDs are estimated with their normalization $N$  defined so that the units of the periodogram are (rms/mean)$^{2}$ Hz$^{-1}$; see Eqn.\ (2) of \citet{2003MNRAS.345.1271V}. In the literature, a power-law model in the form of $P(f) = N~f^{\alpha}$ is usually used to fit the red-noise part of the spectrum $P(f)$ as a function of the frequency $f$,  where  $\alpha \leq 0$ is the power spectral index \citep{1989ARA&A..27..517V} for the PSD of the AGN \citep[e.g.,][and references therein]{2012A&A...544A..80G,2019ApJ...884..125Z,2021ApJ...909..103Z}. The red-noise level (the best fit line to the PSD) is calculated using equations (4--6) of \citet{2005A&A...431..391V}.

\begin{figure*}
    \center
	\includegraphics[width=\linewidth]{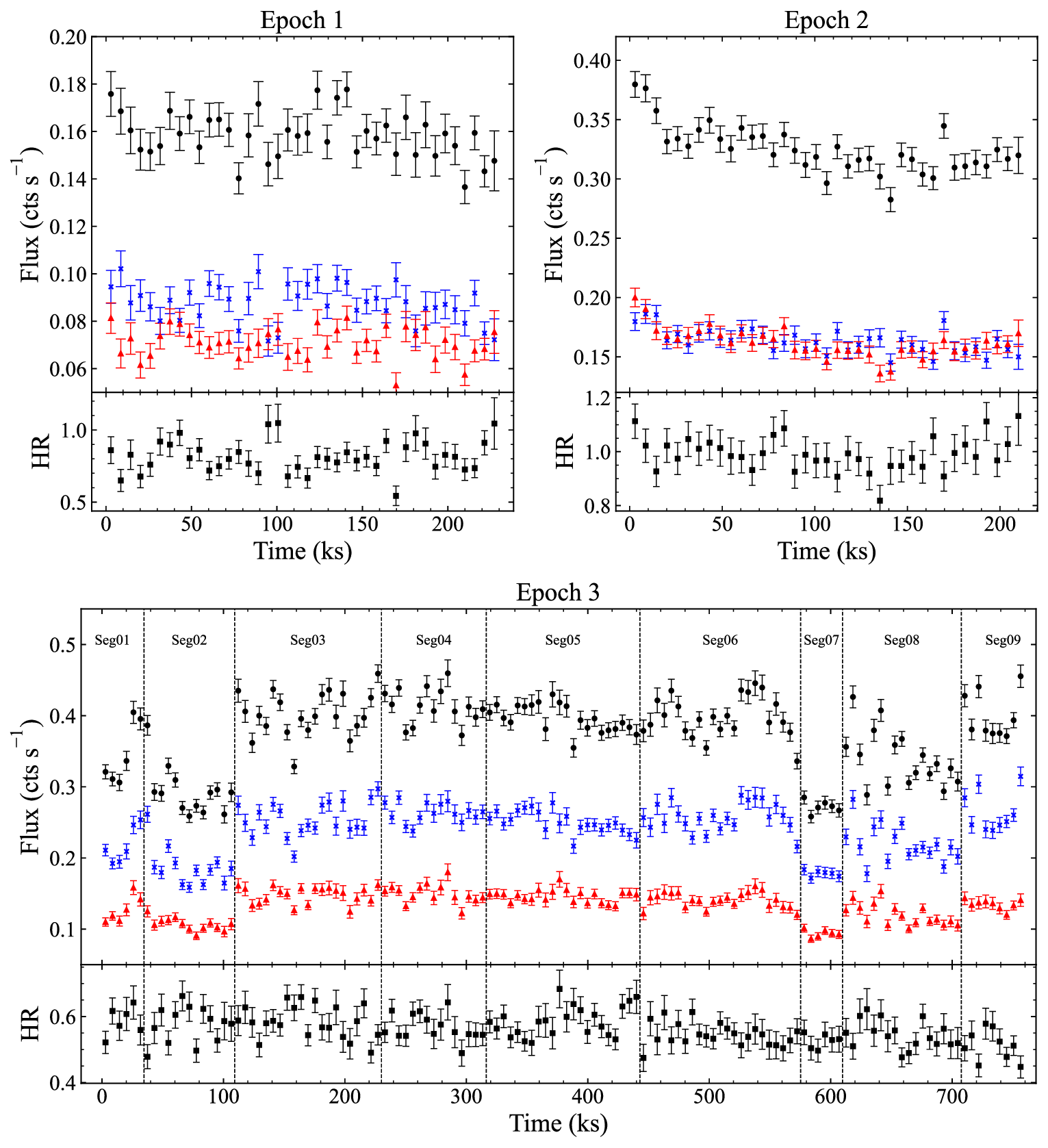}
    \caption{Background-subtracted light curves and hardness ratios (HRs) of the 3 epochs, using the count rates extracted from the source region with the XIS $0+3$ detectors. The light curves,  in 5752-second bins, are: total (0.5 -- 10 keV, black circles); soft 0.5 -- 2 keV, blue crosses); and hard (2 -- 10 keV, red triangles). In the bottom panel, nine segments were designated for studying the spectral variation of epoch 3 in section \ref{sec:excess}.}
    \label{fig:curve}
\end{figure*}

\begin{table}
\caption{X-ray variability parameters}
 \scalebox{0.94}{
	\label{tab:var_par}
 \hspace*{-0.2in}
	\begin{tabular}{ccccc} 
		\hline \hline
              & \multicolumn{3}{c}{ $F_{var}(\%)$}   &  \multicolumn{1}{c} {$\tau_{var}$ (ks)}  \\
        \cline{2-4}
    
        epoch & Soft              & Hard             & Total                  &  Total             \\
              & (0.5 -- 2 keV)   & (2 -- 10 keV)   & (0.5 -- 10 keV)       & (0.5 -- 10 keV)    \\
        \hline         
        1     & ~6.3 $\pm$ 1.4    & ~4.3 $\pm$ 2.0   & ~3.6 $\pm$ 1.2         &	~59.7 $\pm$ 42.5      \\
        2     & ~4.7 $\pm$ 1.0    & ~5.9 $\pm$ 0.8   & ~5.4 $\pm$ 0.6         & ~77.3 $\pm$ 46.9      \\
        3     & 11.3 $\pm$ 0.4    & 12.2 $\pm$ 0.5   & 11.5 $\pm$ 0.3         & 110.8 $\pm$ 13.8  \\
        \hline
        \end{tabular}
        } \\
        \footnotesize
        \textbf{Notes.} $F_{var}$= the fractional rms variability amplitude. $\tau_{var}$= the flux variability timescale.
\end{table}

\section{Results} \label{sec:Results}

\subsection{Intraday Flux Variability} 
We generated LCs, binned at 5752s, of these three epochs of {\it Suzaku} observations of the blazar OJ 287 using three XIS energy data sets (soft, hard, and total), and have plotted them in the upper panels of the sub-figures of Figure \ref{fig:curve}. On visual inspection, it is certainly appears that the LCs of all three epochs in all the energy bands show significant flux variations on IDV timescales. To quantify the IDV flux variability results, we estimated the fractional $rms$ variability amplitude and its error for all the LCs. The presence of IDV is confirmed by the values of $F_{var}$ for the soft, hard and total energy bands during the three epochs that are given in Table \ref{tab:var_par}. We also estimated the flux-normalized IDV timescales and their errors for the XIS total (0.5 -- 10 keV) LCs and the results are provided in the last column of Table \ref{tab:var_par}. \\
\\
In earlier studies based on X-ray flux variability of blazars on IDV timescales, it was found that LSP and ISP blazars show lower amplitude IDV \citep[e.g.,][and references therein]{2015MNRAS.451.1356K,2016MNRAS.462.1508G,2022MNRAS.511.3101P} in comparison to HSP blazars \citep[e.g.,][and references therein]{2017ApJ...841..123P,2018ApJ...859...49P,2018MNRAS.480.4873A,2019ApJ...884..125Z,2021ApJ...909..103Z,2021MNRAS.506.1198D,2022ApJS..262....4N,2022ApJ...939...80D}. Our IDV results for the LSP OJ 287 are thus consistent with those previous results.

\begin{figure*}
    \center
	\includegraphics[width=\linewidth]{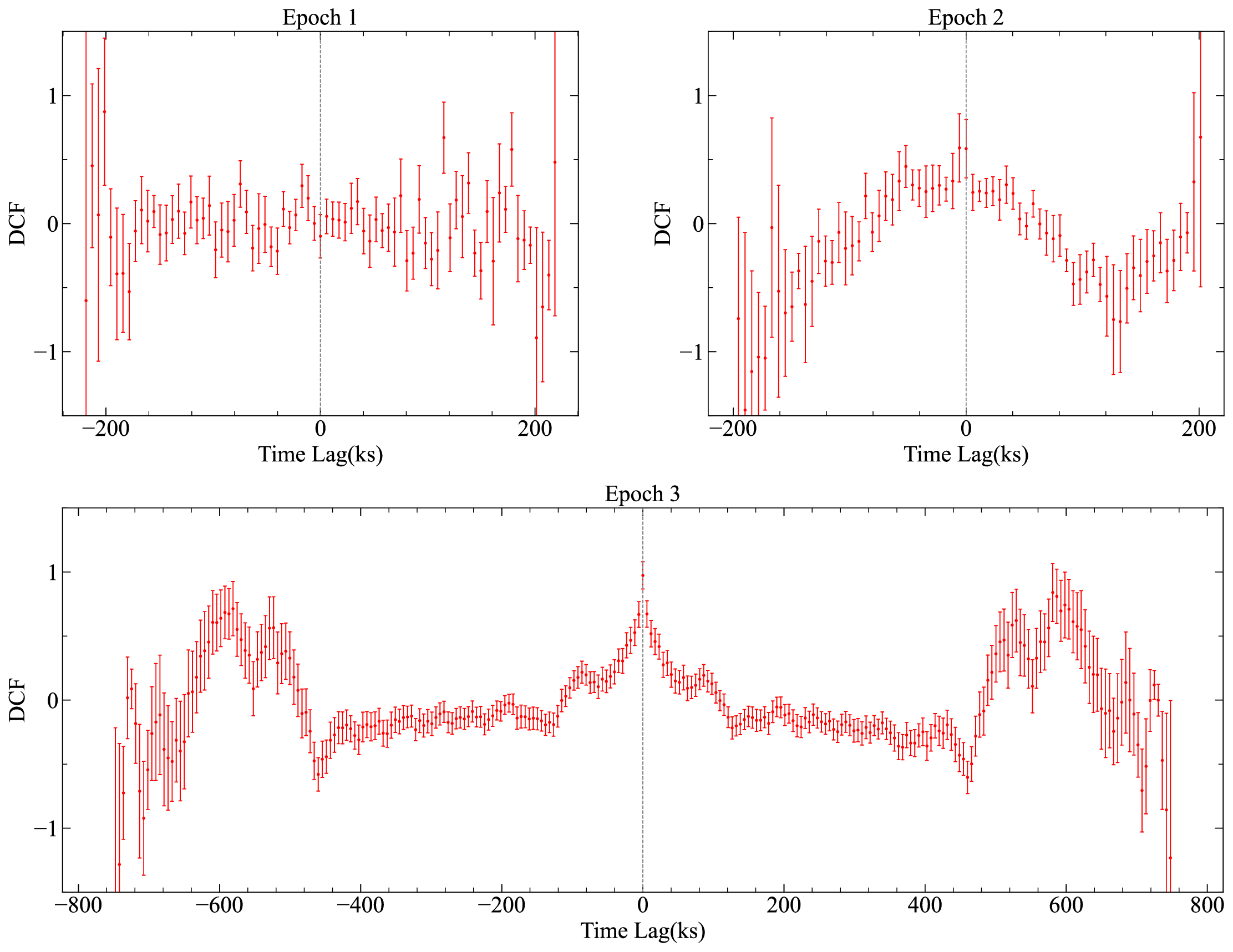}
    \caption{Cross-correlation analysis by DCF for soft (0.5 -- 2 keV) and hard (2 -- 10 keV) for the 3 \Suzaku epochs binned by 5752s.}
    \label{fig:dcf}
\end{figure*}

\subsection{Intraday Cross Correlated Variability}
To investigate if there is a time lag between soft and hard X-ray energies, we carried out DCF analyses as described in Section 3.3 between XIS soft (0.5 -- 2 keV) and XIS hard (2 -- 10 keV) bands for all the three epochs of these {\it Suzaku} observations of OJ 287.  We present these DCF plots in Figure \ref{fig:dcf}.  We fit the plots with a Gaussian function given by Equation \ref{equ:10}, and the fitting parameters are given in Table \ref{tab:dcf}. A negative lag means the soft band precedes the hard band and vice versa. \\
\\
In Figure \ref{fig:dcf} the DCF plot of epoch 1 is flat and consistent with 0 throughout. The DCF plot of epoch 2 indicates that the soft and hard bands are correlated with a small nominal temporal lag that is consistent with a zero lag. On the other hand,  the correlation is strong in epoch 3 at essentially zero lag. The lags for both epochs 2 and 3 can be considered as being zero because the values reported in Table \ref{tab:dcf} are smaller than the size of the interval in which the data were binned. We note that the lack of clear correlations during epoch 1 is likely caused by its relatively low count rates, while the abundance of data in epoch 3 yields the clearest correlations. These DCF results during these observations support the hypotheses that the X-ray emission in the soft and hard energies are cospatial and emitted from the same population of leptons.

\begin{table}
	\centering
	\caption{Correlation Analysis between X-Ray Bands.}
 \scalebox{1.0}{
	\label{tab:dcf}
	\begin{tabular}{ccc} 
		\hline \hline
        epoch        & m(ks)          & $\sigma$(ks) \\ 
         \hline
        1            & ---            & ---     \\
        2            & $-5.41\pm3.4$  & $-23.3\pm6.2$   \\
        3            & $-2.19\pm4.7$  & $-28.4\pm7.8$     \\
        \hline
        \end{tabular}
        } \\
        \footnotesize
        \textbf{Notes.} $m$ = time lag at DCF peaks, $\sigma$ = width of the Gaussian function
\end{table}

\subsection{Intraday Power Spectral Density Analysis}
\begin{figure*}
    \center
	\includegraphics[width=\linewidth]{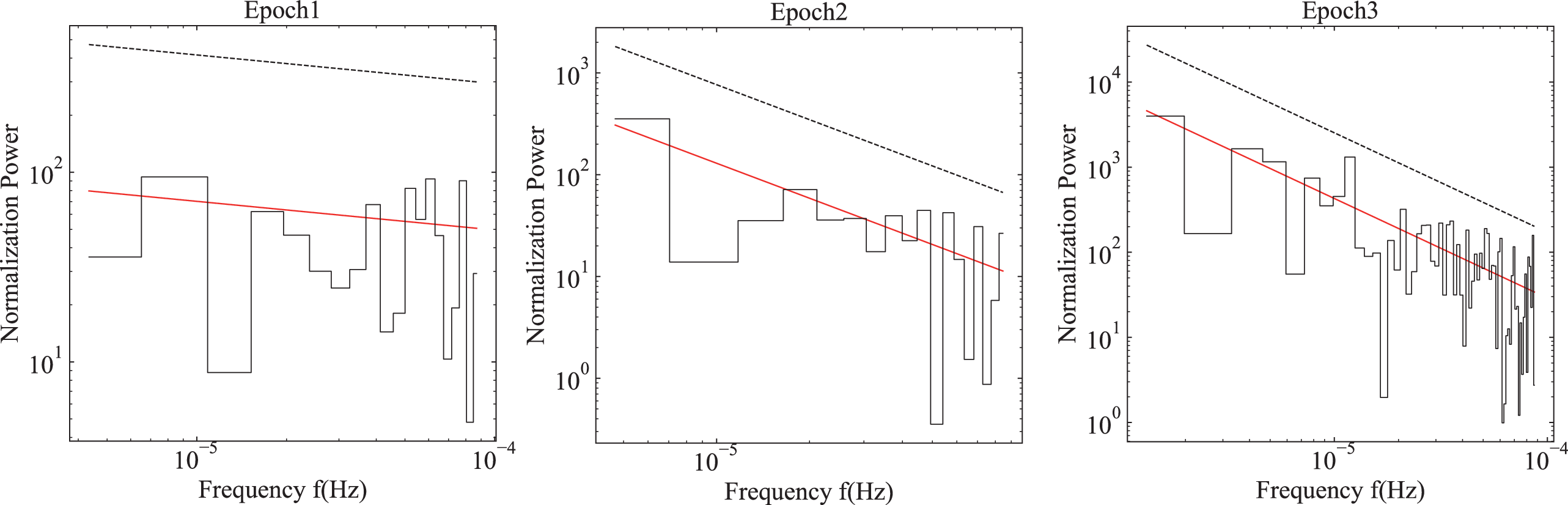}
    \caption{Power spectral densities (PSDs) of the three XIS total (0.5 -- 10 keV) LCs. The continuous red lines are the fitted red-noise and the dotted black line shows the 99.73\% (3$\sigma$) confidence level for any QPO within the red-noise model.}
    \label{fig:psd}
\end{figure*}

We have performed PSD analyses of all the three epochs of the XIS total energy (0.5--10 keV) LCs to characterize the temporal variations of flux and to search for any QPOs in the X-ray emission of OJ 287. These PSD plots are presented in Figure \ref{fig:psd}. We found that the PSDs of all the three epochs can be considered some variant of red noise and there is no detection of any QPO. The slopes, $\alpha$, of the power-law fits to the PSDs, and the logarithmic normalization constants are reported in Table \ref{tab:psdfit}. The average $\alpha = -0.82$. \\
\\
While there seems to be a major difference between the essentially white-noise character of the PSD in epoch 1 and the essentially flicker-noise character of the variations during epochs 2 and 3, we note that the errors on all of these fits are substantial. Further, the lower count rates during epoch 1 make its flat PSD less reliable. The slopes obtained here for the red noise are consistent with those computed for the X-ray fluctuations seen in a wide range of AGNs \citep[e.g.][]{2012A&A...544A..80G}. 

\begin{table}
\centering
\caption{Fitting parameters of PSD in 3 epochs.}
\scalebox{1.0}{
 \label{tab:psdfit}
 \begin{tabular}{ccc} 
  \hline \hline
  epoch    & $\alpha$        & $log(N)$ \\ 
  \hline
  1        & $-0.15\pm1.91$  & $1.09\pm8.55$     \\
  2        & $-1.14\pm2.06$  & $-3.59\pm9.18$   \\
  3        & $-1.17\pm0.93$  & $-3.21\pm4.18$     \\
  \hline
  \end{tabular}
  } \\
  \footnotesize
  \textbf{Notes.} A power-law model is assumed with: $P(f)=Nf^{\alpha}$ and $\alpha<0$.
 \end{table}

\subsection{Intraday Spectral Variability} 
\begin{figure}
    \center
	\includegraphics[width=\linewidth]{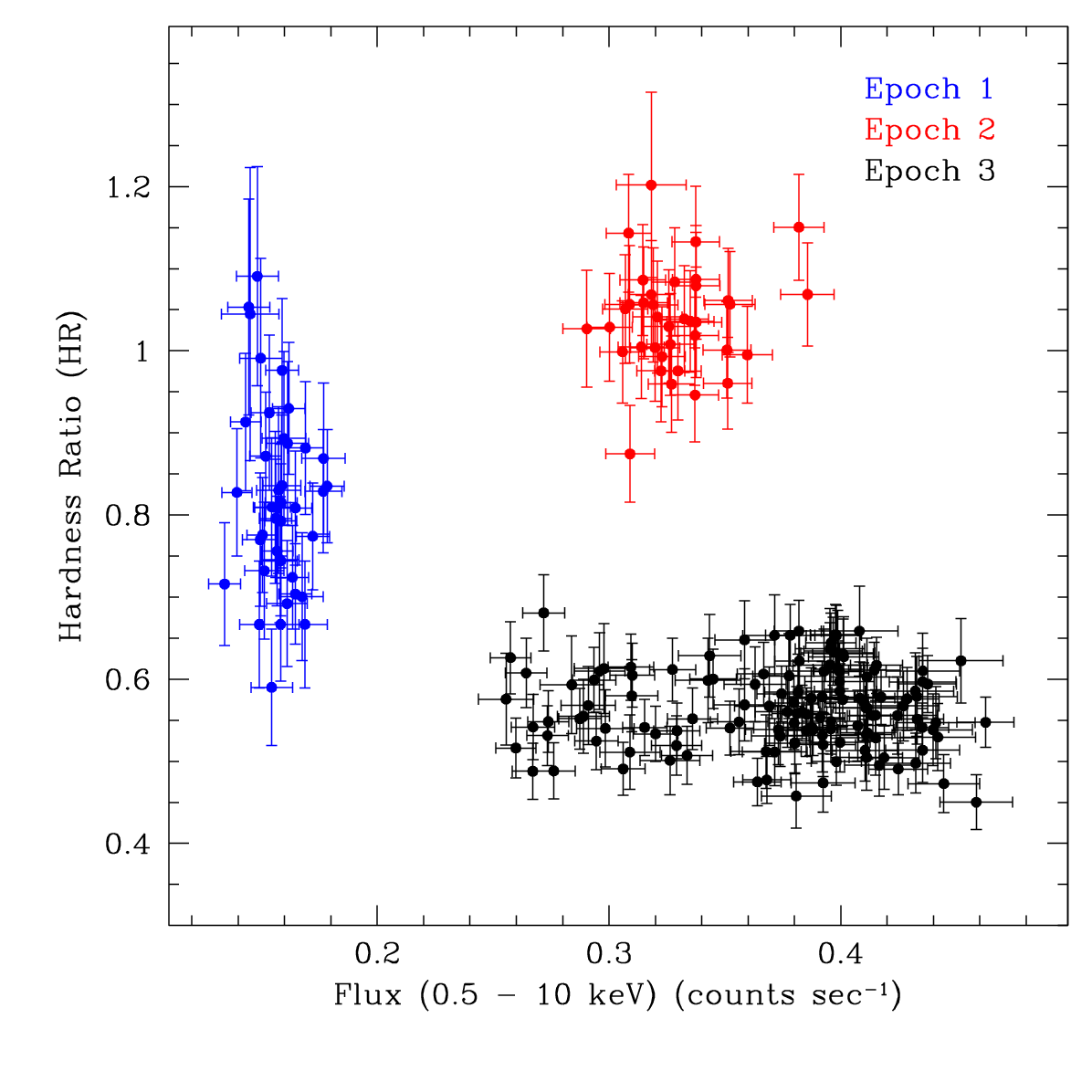}
    \caption{Hardness ratio versus total flux (0.5 -- 10 keV) of OJ 287 in different epochs, based on the time bins in the light curves of Figure \ref{fig:curve}.}
    \label{fig:hardnessflux}
\end{figure}

By using X-ray fluxes in hard and soft energy bands, X-ray spectral variations can be examined using the hardness ratio (HR), defined as
\begin{equation}
    HR=\frac{H}{S}
\end{equation}
where $H$ and $S$ are the net count rates in the hard and soft energy bands, respectively. With $HR$, we can easily characterize any gross spectral variations of the X-ray emission. The error, $\sigma_{HR}$, is calculated from the two errors in hard and soft bands as
\begin{equation}
    \sigma_{HR} = {\frac {2} {(H + S)^2}} \sqrt{(H^{2}\sigma^{2}{_S} + S^{2}\sigma^{2}{_H})} .
\end{equation}
For all three epochs of {\it Suzaku} observations of OJ 287, HR with respect to time is plotted in the bottom panels of sub-figures of Figure \ref{fig:curve}.  We plot the HR  against flux during each epoch in Figure \ref{fig:hardnessflux}. We quantify the variability of the spectra using a standard $\chi^2$ test,
\begin{equation}
    \chi^2 = \sum_{i=1}^{N} \frac{\left(x_{i}-\bar{x}\right)^{2}}{\sigma_{i}^{2}},
\end{equation}
where $x_i$ is the HR value, $\sigma_i$ is the corresponding error, and $\bar{x}$ is mean value of HR. We consider  a significant variability in the HR to be detected if $\chi^2 > \chi^2_{99,\nu}$, where $\nu$ is the number of degrees of freedom (DoF) and the significance level is 0.99. These values are provided in Table \ref{tab:HR} for the three epochs of observations. From Table \ref{tab:var_par} we notice that the variability amplitudes are nearly the same in the soft, hard and total bands in each of the three epochs of observations. From Figure \ref{fig:curve}, HR seems to be at most only slightly variable in all three epochs of observations and this is quantified in Table \ref{tab:HR}. Since OJ 287 is a LBL blazar we might expect a softer when brighter trend \citep[e.g.][]{2021ApJ...909..103Z}. But due to the small amplitudes and similar variations in the soft, hard, and total bands, any such feature is not noticeable in these data, as shown in Fig.\ \ref{fig:hardnessflux}.

\begin{table}
	\centering
	\caption{$\chi^2$ tests of HR variability.}
 \scalebox{1.0}{
	\label{tab:HR}
	\begin{tabular}{lccc} 
		\hline \hline
        epoch    & DoF & $\chi^2$ & $\chi^2_{99}$ \\ 
         \hline
        1        &  39 & 64.82    & 62.43              \\
        2        &  36 & 35.97    & 58.62              \\
        3        & 131 & 205.62   & 171.57              \\
        \hline
	\end{tabular}
 } \\
\end{table}

\subsection{Background-subtracted XIS Spectra of OJ 287}
\label{sec:spectra}
\begin{figure}
    \center
	\includegraphics[width=\linewidth]{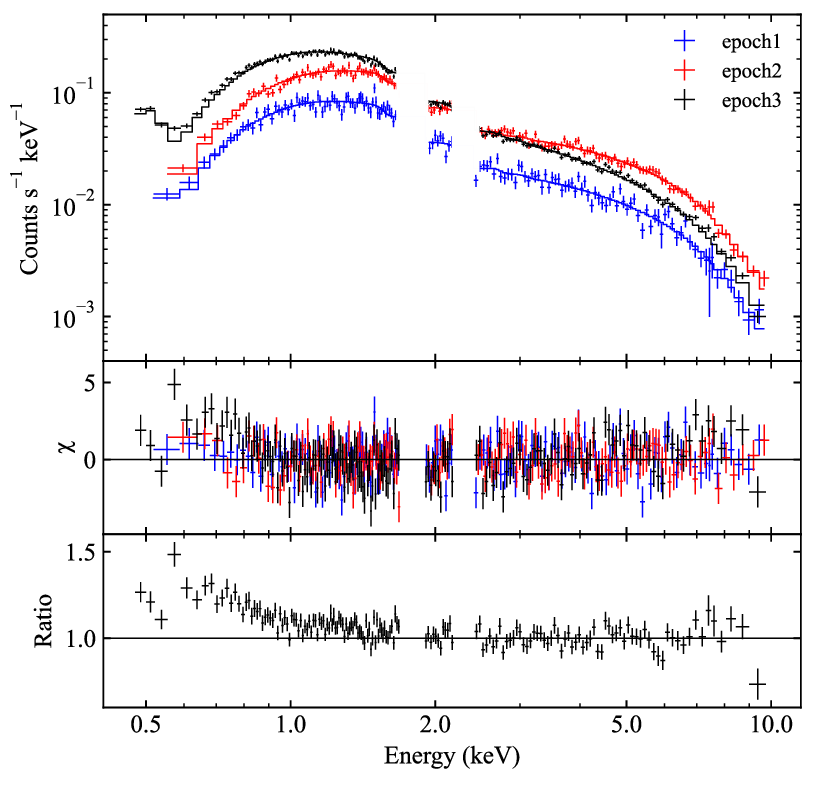}
    \caption{The {\it Suzaku} spectra of OJ 287 in three epochs in 0.5 -- 10 keV from FI CCDs and the best-fitting results by simple absorbed power-law model (top and middle panels). The epoch 1--3 data are shown with blue, red and black points, respectively. Significant residuals are seen below 2 keV and above 7 keV for the spectrum of epoch 3; thus we show the ratio plot of the same fit in 2 -- 10 keV extended down to 0.5 keV for this epoch in the bottem panel.}
    \label{fig:spectra}
\end{figure}

We extracted the spectra of OJ 287 of each observation in three epochs, using the source and background regions described in section \ref{sec:Obeservation}. The counts from the FI chips (XIS 0 and XIS 3) were utilized for all  three epochs, while those from the BI chip (XIS 1)  were also included  for epoch 3. We calculated the response matrix function and auxiliary response files with the HEAsoft commands {\it xisrmfgen} and {\it xissimarfgen}, respectively. The contamination on the XIS optical blocking filters is considered in {\it xissimarfgen} \citep{2007PASJ...59S.113I}. Then we generated a response file by combining the response matrix function and auxiliary response files with the command {\it marfrmf}. In the next step, the source and background spectra of the FI chips for all epochs were combined with the command {\it mathpha}, with the response files added by {\it addrmf}. Then the spectra and response files of the three observations of epoch 3 were further combined by the above methods.\\
\\
We first analysed the  background-subtracted spectra of the FI chips for the three epochs by fitting them with Power-Law (PL) models. Since the FI chips are more sensitive in the high energy bands, we used 0.5--10 keV for the XIS 0+3 spectra. To avoid the calibration uncertainties around the instrumental silicon K-edge and the gold M-edge, we excluded the energy ranges of 1.7--1.9 keV and 2.2--2.4 keV in the spectra, respectively. The model is given by
\begin{equation}
    \centering
    A(E) = KE^{-\Gamma},
\end{equation}
where $K$ is the normalization, and here $\Gamma$ is the photon index. We performed the fitting using XSPEC \citep{1996ASPC..101...17A}, employing $\chi^2$ minimization statistics. Considering the effect of the galactic absorption, the model was modified by the Tuebingen-Boulder interstellar medium absorption model \citep[TBABS;][]{2000ApJ...542..914W} with the hydrogen column density value $N_{H} = 2.56\times10^{20}$ cm$^{-2}$. We got $\alpha$ values of $1.68\pm0.02$, $1.52\pm0.01$ and $2.03\pm0.01$, with $\chi^{2}/\nu$ (where $\nu$ stands for the degree of freedom) of 170.3/149, 153.9/150 and 320.5/168, respectively for epochs 1, 2 and 3, as shown in Figure \ref{fig:spectra}. The spectra were well fitted in epochs 1 and 2, with the results being consistent those of \citet{2009PASJ...61.1011S}. Epoch 3 presented a poor fit by the PL model, as can be seen in the middle panel. By conducting the same fitting in the band 2--10 keV band for this epoch and extending the model to 0.5 keV, a clear soft excess is shown in the bottom panel. We will discuss it in section \ref{sec:excess}. \\
\\
Then we adopted the Broken Power-Law (BPL) model and Log-Parabolic (LP) model to conduct a simultaneous fitting for the spectra of FI and BI CCDs of epoch 3. These two models have been extensively used in X-rays and other emission bands for fitting the curved spectra of a large number of blazars \citep[e.g.,][and references therein]{2002babs.conf...63G,2004A&A...413..489M,2004A&A...422..103M,2006A&A...448..861M,2008A&A...478..395M,2014MNRAS.444.3647B,2017MNRAS.469.3824K,2019ApJ...880...19K,2017ApJ...841..123P,2018ApJ...859...49P,2017ApJ...850..209G,2018MNRAS.473.3638G,2021ApJ...914...46G,2018A&A...619A..93B,2022MNRAS.510.5280M} and even in OJ 287 \citep{2020ApJ...890...47P,2018MNRAS.479.1672K,2022JApA...43...79K}. The BPL model is described as
\begin{equation}
    \centering
    A(E)=\left\{\begin{array}{ll}
    KE^{-\Gamma_{1}} & \text { if } E \geq E_{b}; \\
    KE^{-\Gamma_{2}} & \text { otherwise },
    \end{array}\right.
\end{equation}
where $K$ is the normalization, $E_{b}$ is the spectral break energy, $\Gamma_{1}$ and $\Gamma_{2}$ are the photon indices at high and low energies, respectively.  
The LP model \citep{2004A&A...413..489M} is parameterized with functions of the following form 
\begin{equation}
    \centering
    A(E)= K(E/E_{1})^{-(\alpha-\beta\hspace{0.1cm}log(E/E_{1}))},
\end{equation}
here $A(E)$ is the source flux in units of photons cm$^{-2}$ s$^{-1}$ keV$^{-1}$, $E_{1}$ is the pivot energy, and now $\alpha$ represents the photon index at energy E$_{1}$ and $\beta$ is the spectral curvature parameter. We fixed $E_{1}$ at 1 keV and therefore the LP spectrum is entirely described only by $K$, $\alpha$, and $\beta$. The fitting parameters are presented in Table \ref{tab:specfit}.\\
\\
To choose the model that fits the best, we calculated the F-statistic and its probability of LP and BPL in XSPEC, while PL is taken to be the null hypothesis (NH). If the probability of the NH under an alternative model is under $\leq$ 0.1, or the significance of the alternative model is $\geq90\%$, we choose the model with the highest significance as the best-fit model; otherwise the simple PL model was considered to be the best. The fitting parameters and results of the F-tests are presented in Table \ref{tab:specfit}. Both the LP and BPL models provide better fits than the PL model, especially the BPL. Since other models are considered especially for the soft excess below 2 keV in epoch 3, we will discuss the results more thoroughly in the next section.

\begin{figure*}
    \centering
    \includegraphics[width=1\linewidth]{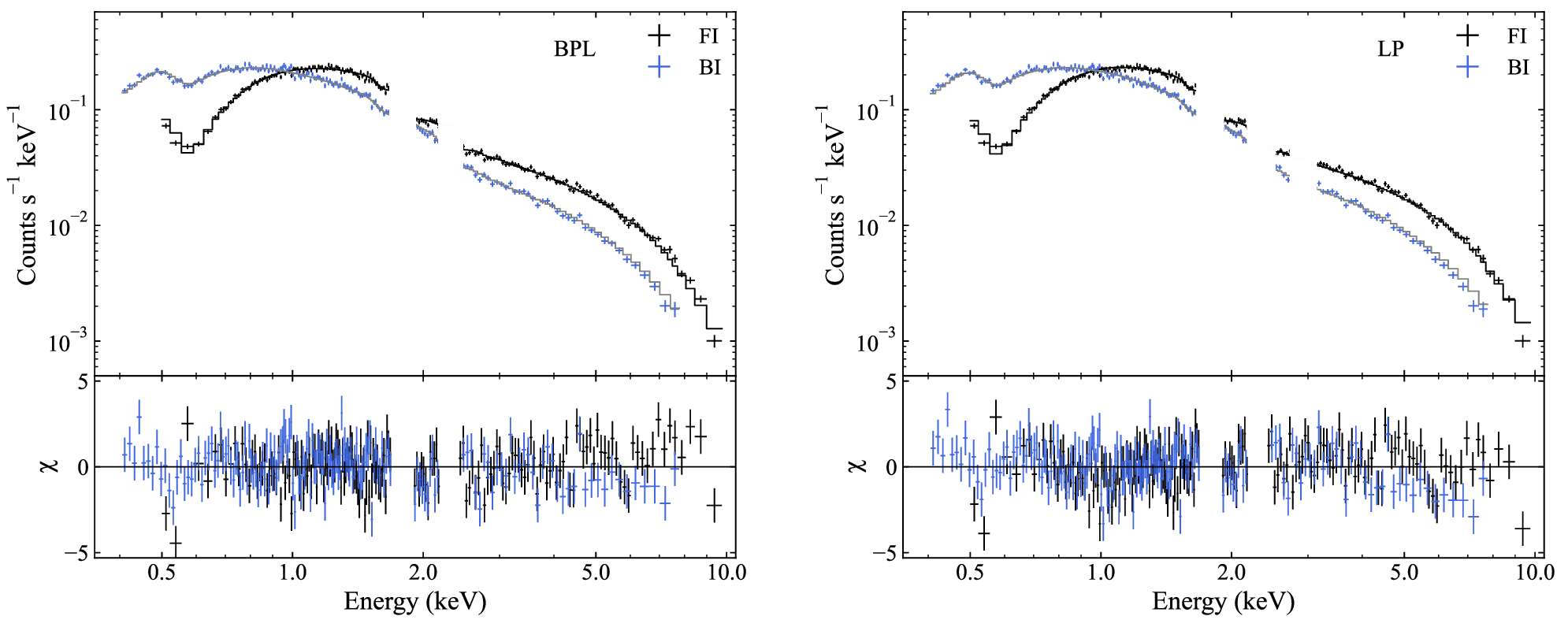}
    \caption{Best simultaneous fitting results for FI (0.5 -- 10 keV) and BI (0.4 -- 8 keV) spectra of epoch 3 and its residuals with a BPL model (left panel) and a LP model (right panel). The FI and BI data are shown with black and blue points, respectively. }
    \label{fig:bpl_lp}
\end{figure*}

\begin{table}
	\centering
	\caption{Parameters of model fitting to the {\it Suzaku} XIS FI+BI spectra of epoch 3.}
    \scalebox{1.0}{
	\label{tab:specfit}
 \hspace*{-0.3in}
	\begin{tabular}{cccc} 
		\hline \hline
        Model             & Component      & Value            \\ 
                          &                &                       \\ 
        \hline
        tbabs             & $N_H(10^{22})$ & 0.0256                \\
                          &                & (frozen)         \\
        \hline
        PL                & $\Gamma$       & 2.07               \\
         &  $\chi^{2}$/d.o.f.              & 774.8/328           \\ 
            &    F-test (prob.)            &  NH               \\ 
        \hline
        LP                & $\alpha$       & 2.18$\pm$0.01        \\                  
                          & $\beta$        & -0.21$\pm$0.01     \\
                          & $E_{p}$(keV)   & 1                 \\
         &      $\chi^{2}$/d.o.f.          & 475.7/327           \\   
            &    F-test (prob.)            &  205.6 ($<$ 0.01)  \\ 
        \hline
        BPL               & $\Gamma_{1}$   & 2.31$\pm$0.02     \\
                          & $\Gamma_{2}$   & 2.01$\pm$0.02       \\
                          & $E_{b}$(keV)   & 1.08$\pm$0.02     \\
         &     $\chi^{2}$/d.o.f.           & 453.8/326         \\  
            &      F-test (prob.)          &  115.3 ($<$ 0.01)  \\ 
        \hline
        \end{tabular}
        }\\
        \noindent
        \footnotesize
        \textbf{Notes.} Col.1: models used in our fitting; Col.2: components for each model; Col.3: the fitting parameters of each model component. 
	\end{table}

\subsection{The soft excess of the X-ray spectrum in epoch 3}
\label{sec:excess}

\begin{figure*}
    \center
	\includegraphics[width=\linewidth]{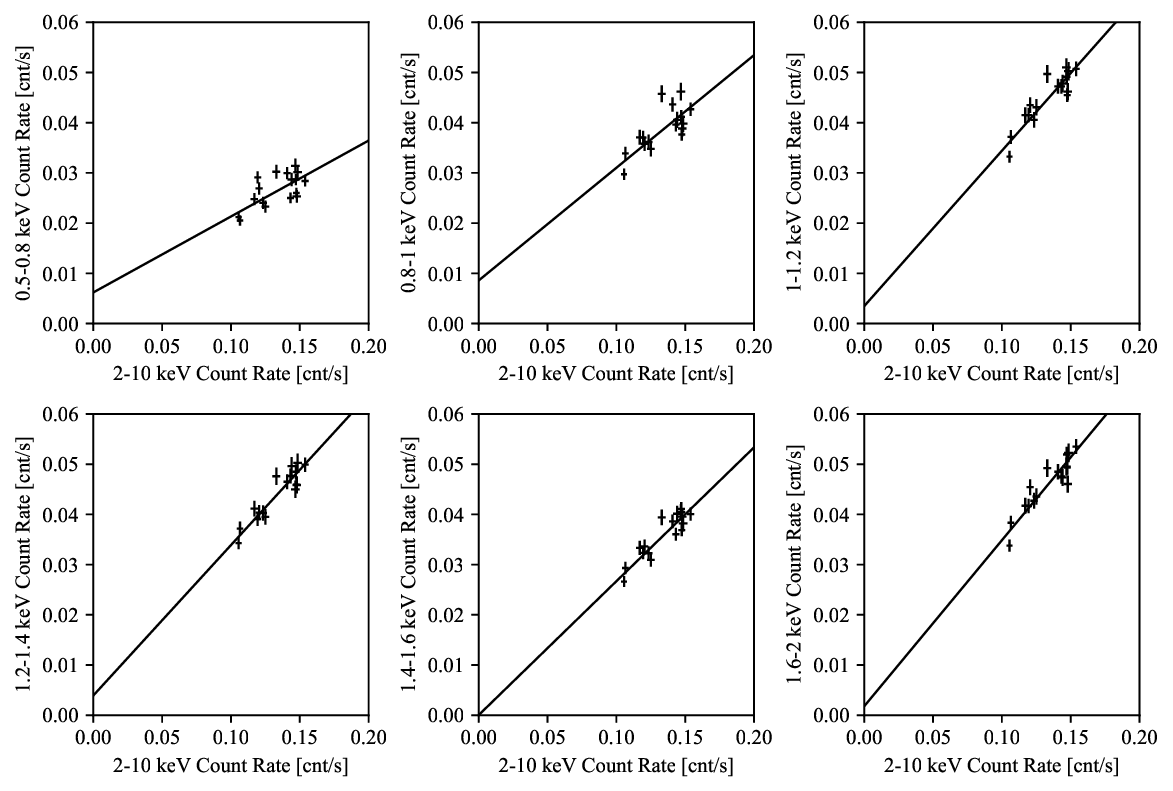}
    \caption{Count Count Plots of epoch 3 and the linear fits to them. The abscissas give background-subtracted FI count rates in 2 -- 10 keV, while the ordinates represent the background-subtracted FI count rate of six finer soft energy bands: 0.5 -- 8 keV, 0.8 -- 1 keV, 1 -- 1.2 keV, 1.2 -- 1.4 keV, 1.4 -- 1.6 keV, 1.6 -- 2 keV.}
    \label{fig:ccp}
\end{figure*}

\begin{figure}
    \center
	\includegraphics[width=\linewidth]{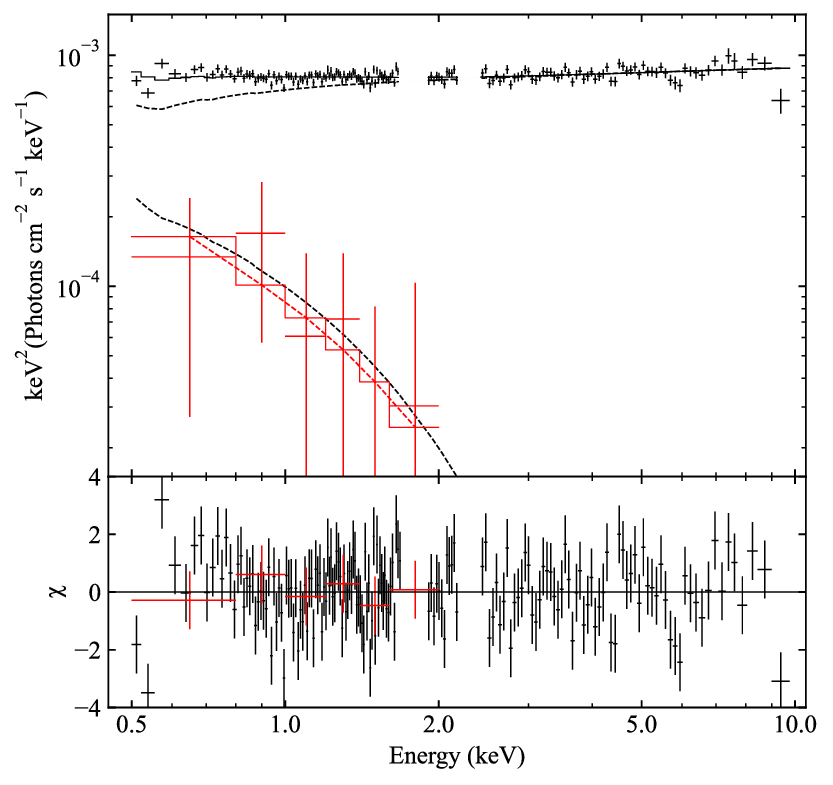}
    \caption{The best simultaneous fitting results of epoch 3 spectra obtained from FI (black points, S2) and the soft component using the values of $b$ in Table \ref{tab:CCPs_fitting} convolved by the response function of FI (red crosses, S1). The dashed lines are components from the fitting by the model \texttt{optxagnf+PL} with the parameters shown in Table \ref{tab:optxagn}.}
    \label{fig:optxagnf}
\end{figure}

In epoch 3 of our \Suzaku observation a soft excess clearly is present below 2 keV, as is seen by fitting the source's FI spectra in 2--10 keV using a PL model (Section \ref{sec:spectra}). This  feature of OJ 287 was first reported in an \textit{ASCA} observation in November 1994 and attributed to a ``synchrotron soft tail" \citep{2001PASJ...53...79I}. Recently the same feature was seen in an \textit{XMM-Newton} observation in May 2015 (MJD = 57149--57150), and was proposed to be explained in two ways, either in terms of cool Comptonization in the accretion disc, or blurred reflection from the partially ionized accretion disc, by \citet{2020ApJ...890...47P}. They suggested that the latter was favored based on the lag of the UV emission with respect to the X-rays. In that work, model degeneracy is a problem even with multi-band spectra inducing optical UV data. 
Fortunately, the \Suzaku observation of epoch 3 includes the entire period of this \textit{XMM-Newton} observation but spans five times more elapsed time, allowing us to investigate this structure in more detail.\\
\\
We now explore a model independent method to extract the soft excess component from the the entirety of the PL dominated continuum. In the works of \citet{2011PASJ...63S.925N,2013PASJ...65....4N}, a method called Count–Count Correlation with Positive Offset (hereafter C3PO) was provided for relatively long observations exhibiting significant variations in, and correlations between, different energy bands in the light curves. This is similar to what we can see in the bottom panel of Figure \ref{fig:curve}. To further explore the correlations between the soft and hard bands, we divided the 0.5 -- 2 keV band into six finer bands: 0.5 -- 0.8 keV, 0.8 -- 1 keV, 1 -- 1.2 keV, 1.2 -- 1.4 keV, 1.4 -- 1.6 keV and 1.6 -- 2 keV, and plotted the correlation of the count rates in these finer bands with that in the 2 -- 10 keV hard band for the whole observation period of epoch 3, as shown in six Count-Count Plots (CCPs) in Figure \ref{fig:ccp}. In each CCP, the data can be fitted by a linear expressions 
    $y = ax + b$,
indicating strong correlations between soft and hard bands through the positive slopes $a$. As shown in Table \ref{tab:CCPs_fitting}, the values of the intercept parameter $b$ all exhibit nominally positive offsets, indicating the presence of ``more stable" components which are independent of the hard band variations.\\
\\
Next, we convolved our fitting results for parameter $b$ by the response function of the \Suzaku FI CCDs to produce the spectrum of the stable soft component, plotted in red crosses in Figure \ref{fig:optxagnf}. To investigate the properties of this component, we simultaneously fit its spectrum together with the FI spectrum of the whole epoch 3, using the \texttt{optxagnf} model \citep{2012MNRAS.420.1848D}, similarly to what was done in \citet{2020ApJ...890...47P}. In this scenario, the gravitational potential energy is released as blackbody emission beyond the radius of $R_{\rm{corona}}$. Within this coronal radius,  optically thick, low temperature thermal Comptonisation produces the soft X-ray excess while the optically thin, high temperature thermal Comptonisation produces the power-law emission which dominates above 2 keV. We fixed the central black hole mass at $2\times10^{10}M_{\sun}$ \citep{2016ApJ...819L..37V,2018MNRAS.473.1145K}, and the source luminosity distance $D_{\rm L}$ at 1677 Mpc. To show the power-law component intuitively, we set the fraction of the power below $R_{\rm{corona}}$ of $f_{\rm pl} = 0$, and  added a PL model to \texttt{optxagnf}. The fitting resulted in $\chi^{2}$/$\nu = 228/167$, with the details of the best-fit parameters listed in Table \ref{tab:optxagn}. We noticed that the fitting statistics are much better than those of the PL, LP, or BPL models in section \ref{sec:spectra}. Comparing to the PL model as the null hypothesis, the F-test results in a value of 11.1, and 31.1 (probabilities both < 0.01) by adding the BI data, giving the best solution so far when compared to the LP and BPL models. Moreover, the resulting parameters are quite consistent with those derived by \citet{2020ApJ...890...47P}, in which multi-band data where used, including \textit{XMM-Newton} observations, to further support the suitability of our C3PO analysis. \\
\\
We calculated the fluxes of different components in the joint spectral fitting and show them in Table \ref{tab:optxagn}. In 0.5 -- 2 keV the flux of the soft stable component extracted from CCP (S1) is at the level of 90\% of the flux of the cool Comptonisation component from S2 (also shown in Figure \ref{fig:optxagnf}), indicating their very similar origins. The latter, however, comprises only 13.4\% of the whole of the flux in 0.5 -- 2 keV, with the bulk of the  emission contributed by the PL model, which is co-variate with the hard emission in 2 -- 10 keV. For the whole of the emission between 0.5 -- 10 keV, the soft excess component  only comprises $\sim$ 6\%, and hence is very difficult to identify solely from the X-ray data, unless it presents  very different variation properties. These can be segregated by the C3PO method as in our study, providing  a strong assist to the multi-band spectral fitting as performed in \citet{2020ApJ...890...47P}.  \\
\\

\begin{table}
	\centering
	\caption{Parameters obtained by fitting CCPs.}
	\label{tab:CCPs_fitting}
	\begin{tabular}{ccc} 
		\hline
        Energy range(keV)  & a   & b       \\ 
        \hline 
        0.5-0.8            & 0.15$\pm$0.04 & 0.0062$\pm$0.0049               \\
        0.8-1              & 0.22$\pm$0.04 & 0.0086$\pm$0.0057              \\
        1-1.2              & 0.31$\pm$0.03 & 0.0035$\pm$0.0045            \\
        1.2-1.4            & 0.30$\pm$0.03 & 0.0039$\pm$0.0036             \\
        1.4-1.6            & 0.27$\pm$0.03 & 4.0$\times10^{-5}$            \\
        1.6-2              & 0.33$\pm$0.03 & 0.0018$\pm$0.0043               \\
        \hline
	\end{tabular}
\end{table}

\begin{table}
	\caption{Simultaneous fitting of the CCP extracted soft component (S1) and the whole spectra (S2) of epoch 3 based on the FI data.}
    \scalebox{1.0}{
	\label{tab:optxagn}
 \hspace*{-0.3in}
	\begin{tabular}{cccc|c} 
		\hline \hline
        Model component   &  Parameter     & S1   &     S2       \\ 
 \hline
        optxagnf          & Acc. rate $\left(\frac{L}{L_{\text {edd }}}\right)$ & $0.044^{+0.02}_{-0.03}$  & --\\
                          & Spin(a)                 & $0.98^{+0.01}_{-0.02}$  & --   \\
                          & Coronal radius($r_{g}$) & 11.66$\pm$2.4           & --   \\
                          & Plasma temp. (keV)      & 0.48$\pm$0.01           &  --  \\
                          & Optical depth ($\tau$)  & 8.78$\pm$0.05           &   --   \\
                          &  PL fraction$f_{\rm pl}$ &  0 (fixed)              & -- \\
                          &  Norm.                  & 0.41$\pm$0.2            &  0.48$\pm$0.03 \\ 
                          &  Flux (0.5--2 keV)     &  2.16                   & 2.40 \\

           PL             & Photon index $\Gamma$  & -- & 1.96$\pm$0.01   \\
                          & Norm.($10^{-3}$)       & 0(fixed)   & 0.76$\pm$0.01       \\
                          & Flux (0.5--2 keV)      & -- & 15.49                \\
                          & Flux (2--10 keV)     & -- & 21.51                \\            
                          & $\chi^{2}$/d.o.f.     & 0.78/5       &   227.2/162      \\   
        \hline
        \end{tabular}
        }\\
        \noindent
        \footnotesize   
        \textbf{Notes.} The flux of each component is in the unit of $10^{-13}$ erg cm$^{-2}$ s$^{-1}$. For S2 the parameters of \texttt{optxagnf} (except for the normalization) are all fixed to S1.
	\end{table}

\section{Discussion}
\label{sec:discussion}
The three \Suzaku observations of OJ 287 presented here were part of coordinated X-ray observations around the  $\sim12$-yr recurrent optical outbursts that are the basis of the claim of the pres
ence of a binary SMBH. The observations can best be understood astrophysically in terms of the phases of the binary orbit of the secondary  SMBH with respect to the accretion disc of the primary during these three epochs \citep[available in][]{2024ApJ...968L..17V}. During epoch 1, the impact on the disc had not yet happened, but was due to come a few months later, so no significant activity in the optical region was expected or seen. During epoch 2, the impact had happened a few months earlier, resulting in a higher accretion rate onto the primary SMBH and thus greater outflow through its jet and an associated increase of optical brightness \citep{1997ApJ...484..180S,2009ApJ...698..781V}. During epoch 3, a disc impact had happened nearly two years earlier, so no increase in optical brightness was expected or seen \citep{2024ApJ...968L..17V}.\\ 
\\
We find that the source is temporally variable on all occasions and is also accompanied by significant spectral changes, as displayed in Figure \ref{fig:spectra}. The flaring X-ray spectrum (epoch 2) is harder compared to the non-flaring (epochs 1 and 3). Of these three epochs the 2015 observation is most strongly variable and the spectral index is also quite different -- intermediate between the source hard and reported super-soft state \citep[e.g.][]{2022MNRAS.509.2696S,2022JApA...43...79K}. Our study shows intra-day X-ray variability, of similar level (Table \ref{tab:var_par}) in the hard and soft bands, in each of the three epochs, and the variations are simultaneous in both the bands, at least for the high flux states (epochs 2 and 3; see Fig. \ref{fig:dcf}). For epoch 1, the DCF values are almost consistent with zero, thanks to its low count rate and modest variability. The brightest of these, epoch 3 in 2015 was the most variable of the three epochs. 
The two 2007 observations have been explored by \citet{2009PASJ...61.1011S} and our respective results (for hardness, spectral model and parameters) are consistent with theirs. We have additionally also explored cross-correlations and PSDs of the flux variability (see Table \ref{tab:psdfit}).\\
\\
The spectra of epochs 1 and 2 are well fitted by the PL model as was also noted in \citet{2009PASJ...61.1011S}, indicating non-thermal jet emissions. For the epoch 3, both LP and BPL models are preferred without combing optical and UV data. The non-thermal spectra combined with IDV and simultaneous variations indicate that the X-ray emission could be primarily jet emission also for this epoch. A flatter X-ray spectrum ($\rm\Gamma_X \sim 2$) is presumably due to the significant contribution of the synchrotron high energy tail in the soft X-ray band which probably caused the soft excess seen in Figure \ref{fig:spectra}. This also makes the count rates of epoch 3 similar to the 2007 flaring episode (epoch 2). However, since the extension of this non-thermal modeling did not fit well with the optical and UV data, another explanation based on a possible accretion disc component, modeled by \texttt{optxagnf} \citep{2012MNRAS.420.1848D} was proposed by \citet{2020ApJ...890...47P}. The XMM-Newton observation, used by them as the evidence for this, overlaps partly with the much longer \Suzaku observation of 2015. Our analysis of the \Suzaku data exploits a new model independent methodology, C3PO \citep{2011PASJ...63S.925N,2013PASJ...65....4N}, illustrated in Figure \ref{fig:ccp}. The convolved spectrum indicates that the origin of the soft X-ray excess below 2 keV (which takes up a flux fraction of $\sim13\%$ in 0.5 -- 2 keV) is similar to the one reported in \citet{2020ApJ...890...47P}, and has a significant contribution from optically thick low-temperature thermal disc Comptonisation. The majority of the whole emission is still from a PL component; although it is computed in terms of the optically thin disc Comptonisation in \texttt{optxagnf}, we cannot exclude the existence of the jet emission in this component because of its simple content and free normalization in the fitting. In the following sections we will discuss both the jet emissions as the conservative primary composition of the X-ray fluxes of this source, and the possible disc emission evidenced as the soft excess in the epoch 3.\\

\subsection{Synchrotron non-thermal jet emissions}
\label{sec:jet}

 OJ 287 is a LSP blazar with the synchrotron portion of its SED peaking around NIR energies \citep{2009PASJ...61.1011S, 2010ApJ...716...30A,2013MNRAS.433.2380K}. Studies of multi-wavelength flux and spectral properties in various sub-classes of blazars seem to require systematic changes of intrinsic physical parameters such as jet size, magnetic field, maximum electron energy, and Doppler factor \citep[e.g.][]{1996ApJ...463..444S}. It has been claimed that LSP blazars have smaller magnetic fields/electron energies and larger sizes as compared to HSP blazars \citep{1996ApJ...463..444S}. In OJ 287, with the synchrotron  peak falling in NIR energies, the X-ray energy measurements (here, 0.5 -- 10 keV by {\it Suzaku}) do not follow the optical-UV spectrum in general \citep[e.g.][]{2009PASJ...61.1011S, 2013MNRAS.433.2380K} and hence, this emission is not purely synchrotron but indicates an intermediate spectral state (with $\rm \Gamma_X \sim 2$). In this state, the synchrotron component contributes substantially to the lower end of the X-ray spectrum, driving the transition from harder to a softer spectrum \citep{2022MNRAS.509.2696S}. The 2015 spectrum is of this type and optical to X-ray spectral studies show that the synchrotron component extends well into the X-ray \citep{2020ApJ...890...47P}.  Hence, the lowest energy portion of the X-ray band provided by {\it Suzaku} used in the present study lies on the high-frequency tail of the synchrotron emission. In other studies of OJ 287, the X-ray fluxes did not follow the optical-UV flux increases in the December 2015 flare \citep{2016ApJ...819L..37V} while the opposite was true in the 2020 flare \citep{2020MNRAS.498L..35K}. These distinct behaviours are expected if the former is an impact flare and the latter a tidal flare \citep{1997ApJ...484..180S}.\\
\\
Therefore, synchrotron losses can be used to get an estimate/constraint on the magnetic field \citep{2020ApJ...890...47P}. By adopting the simplest causality argument for the X-ray emission from a relativistic jet we can use the minimum variability timescale $\tau_{var,min}$ from Eqn.\ (1) to estimate the upper limit to the size of the emitting region, $R$ \citep[e.g.][]{2021ApJ...909..103Z}
\begin{equation}
R \leq {\frac{\delta}{1+z}}c\tau_{var,min}.    
\end{equation}
The value of the Doppler factor, $\delta$, for OJ 287 is found in the range of 3.4 $-$ 18.6 \citep[e.g.][]{1969ApJ...155L..71K,2022MNRAS.509.2696S} from a variety of methods involving different bands. Using the minimum variability timescale of 110.8 ks (Table \ref{tab:var_par}) and this full range of $\delta$, we estimate the size of the emitting region to be in the range $\rm \sim 8.6 \times {10}^{15} - 4.7 \times {10}^{16}$ cm, in line with the range found in SED modeling using optical and gamma-ray variability \citep[e.g.][]{2022MNRAS.509.2696S,2013MNRAS.433.2380K}. This too supports a dominant non-thermal jet origin for this component. \\
\\
Following  \citet{2019ApJ...884..125Z,2021ApJ...909..103Z} we estimate the synchrotron cooling timescale, t$_{cool}(\gamma)$ in the observed frame  of a relativistic electron with ${\it E = \gamma m_{e}c^{2}}$ \citep[see, e.g.,][]{1979rpa..book.....R} 
\begin{equation}
t_{cool}(\gamma) \simeq 7.74\times10^{8}\frac{(1+z)}{\delta} B^{-2}\gamma^{-1} ~{\rm s}.
\end{equation}
Here $B$ is  the magnetic field in Gauss. For these {\it Suzaku} observations, the observed synchrotron emission frequency is 
\begin{equation}
\nu \simeq 4.2 \times 10^{6} {\frac{\delta}{1+z}} B \gamma^{2} Hz \simeq 10^{18} \nu_{18} ~{\rm Hz}.
\end{equation}
Here 0.12 $\leq \nu_{18} \leq$ 2.42 for X-rays 
spanning {\it Suzaku}’s total employed energy range of 0.5 -- 10 keV. The cooling timescale should be longer than or equivalent to this shortest variability timescale \citep[e.g.][]{2015ApJ...811..143P}
\begin{equation}
 t_{cool}(\gamma) \leq \tau_{var,min}.   
\end{equation}
By combining equations (11) and (12), we find an equation for $t_{cool}(\gamma)$ which is independent of $\gamma$, and then we substitute for it in equation (13) with the minimum variability timescale ($\tau_{var,min}$) of 110.8 ks  we found for OJ 287 using $F_{var}$ $\simeq$  11.3\% (Table \ref{tab:var_par}) in the soft X-ray energy range of 0.5 -- 2 keV. We thus find a bound on the magnetic field of this emitting region of OJ 287 to be 
\begin{equation}
B \geq 0.06 \ \delta^{-1/3}\nu^{-1/3}_{18} ~{\rm G}.
\end{equation}
For the full range of $\delta =$ 3.4 -- 18.6 reported in the literature \citep{1969ApJ...155L..71K,2022MNRAS.509.2696S} for OJ 287, we get a a range for the lower limit to the  magnetic field, $B > (0.02 - 0.04) \ \nu^{-1/3}_{18}$ G. Earlier estimations of the magnetic field
for OJ 287 at different flux states and different epochs
vary between 0.7 and 11.5 G \citep[e.g.,][]{2018MNRAS.473.1145K,2018MNRAS.479.1672K,2021A&A...654A..38P,2022MNRAS.509.2696S}. As $\nu_{18}$ can be any value in the range of 0.12 to 2.42, our results are consistent with these earlier studies.
From equations (11) and (12), we find the electron Lorentz factor $\gamma$ for OJ 287 through
\begin{equation}
\gamma \leq 2.2 \times 10^{6} \ \delta^{-1/3} \ \nu^{2/3}_{18}.
\end{equation}
Taking the same full range of $\delta$ into consideration, we estimate the upper limit of $\gamma$ to lie between 8.3 $\times \ \rm{10}^{5} \nu_{18}^{2/3}$ to 1.47 $\times \ \rm{10}^{6} \nu_{18}^{2/3}$. Our result is consistent with those derived from broadband SED modeling studies \citep[e.g.,][]{2018MNRAS.473.1145K,2018MNRAS.479.1672K,2021A&A...654A..38P,2022MNRAS.509.2696S}.

\subsection{The soft excess correlated to the accretion disc}
\label{sec:disc}

\begin{table*}
    \centering
    \caption{Fitting results of the FI spectra of nine segments in epoch 3 by the \texttt{optxagnf+PL} model.}
    \label{tab:norm}
    \begin{tabular}{cccccccccc}
    \hline \hline
    Norm.                   & seg.1                & seg.2         & seg.3         & seg.4         & seg.5         & seg.6          & seg.7                & seg.8         & seg.9\\
    \hline
     Norm.(optxagnf)        & 0.65$^{+0.32}_{-0.34}$ & 0.62$\pm$0.2  & 0.36$\pm$0.2  & 0.45$\pm$0.25 & 0.28$\pm$0.19 & 0.42$\pm$0.21  & 0.42$^{+0.28}_{-0.31}$ & 0.62$\pm$0.22 & 0.63$\pm$0.38\\
    \hline 
    Norm.(PL)($10^{-3}$)    & 0.62$\pm$0.07        & 0.52$\pm$0.04 & 0.81$\pm$0.04 & 0.83$\pm$0.05 & 0.82$\pm$0.04 &  0.82$\pm$0.04 & 0.54$\pm$0.06        & 0.66$\pm$0.05 & 0.82$\pm$0.08\\
    \hline
    Photon index $\Gamma$   & 1.85$\pm$0.1         & 1.84$\pm$0.06 & 1.92$\pm$0.04 & 1.93$\pm$0.05 & 1.95$\pm$0.04 & 1.98$\pm$0.05  & 1.95$\pm$0.09        & 1.95$\pm$0.06 & 2.02$\pm$0.08\\
    \hline 
    $\chi^{2}$/d.o.f.       & 90.4/103             & 210.4/215     & 434.8/400     & 313.9/296     & 472.2/463     & 417.8/396      & 125.2/118            & 275.2/265     & 179/165     \\
    \hline
    \end{tabular}
    \label{tab:my_label}
    \\
\end{table*}

A soft excess, where more emission is seen below $\sim$2 keV than would be predicted by extrapolating the power-law spectrum observed at higher energies \citep[e.g.][]{2006MNRAS.365.1067C}, is a prominent feature of the spectra of many AGN in the X-ray bands and was first identified by \citet{1985MNRAS.217..105A}. However, while this phenomenon is not common in blazars, it can be explained as the result of a non-thermal corona model \citep{2008A&A...479..365P} or a blackbody-like component with temperature $\sim$ 0.1 keV \citep{2018MNRAS.473L..89K}. A number of other AGN soft excess models have been proposed, including a slim accretion disc where photon trapping raises the temperature \citep{1988ApJ...332..646A,2000PASJ...52..499M}; in this scenario, the accretion is super-Eddington \citep{2005gbha.conf..290T}. Alternatively, extreme ultraviolet photons from the  accretion disc emission can be Comptonised \citep{2004A&A...422...85P}. This could involve strong, relativistically blurred absorption from a disc wind \citep{2004MNRAS.349L...7G} or photoionized emission blurred relativistically by motion in an accretion disc \citep{2001MNRAS.323..506B,2006MNRAS.365.1067C}.\\
\\
In section \ref{sec:excess} we found a clear soft excess in epoch 3 in 2015, which was also reported in \citet{2020ApJ...890...47P} for the shorter {\it XMM-Newton} observation made during the \Suzaku observation. In earlier studies, a similar spectral state of OJ 287 was first reported by \citet{1997PASJ...49..631I}. \citet{2001PASJ...53...79I} reanalyzed the multi-band data and drew the conclusion that the synchrotron component contributed the soft excess; this is  supported by our study in that both LP and BPL models gave good phenomenological fits. However, a NIR-optical/UV break was found in \citet{2020ApJ...890...47P}, and both the LP and BPL models are inconsistent with the SED fitting results in this period. The extrapolation of the synchrotron spectrum's high energy end cannot match the NIR and optical/UV data at the same time, which may indicate an additional component in the observation. In \citet{2020ApJ...890...47P} the models of \texttt{optxagnf} and \texttt{relxill} were suggested as indications of the existence of the accretion disc component. The latter model, which is the blurred reflection from a hotter, optically thin portion of the disc was favoured by them. However, we consider this possibility is less likely, given no evidence of a broad Fe-K$\alpha$ line near 6 keV \citep[e.g.][]{1995Natur.375..659T}. \\
\\
\citet{2018MNRAS.473.1145K} analyzed the multi-band data of OJ 287 during its impact flare between 2015 November and 2016 May, and for the first time found two bumps in the NIR-optical-UV band. The “big blue bump” in the NIR-optical could be attributed to the multicolor emission of the accretion disc that is related to the primary black hole and appears to be present since 2013 May in the post-impact 
state of OJ 287. The “little blue bump” in the optical-UV may be consistent with the line emission above the bremsstrahlung emission which caused the break between the optical and UV \citep{2012MNRAS.427...77V}. The blue bumps had not been discussed before, and there is no clear evidence that they are connected with the jet emission \citep{2012MNRAS.427...77V}; hence a thermal disc component would presumably be present here and may even be related to the soft excess in X-rays, although the soft X-ray excess is usually present in Seyfert 1 AGN \citep{2004MNRAS.349L...7G,2006MNRAS.365.1067C}. Considering the jet emission to still be the main contribution to the high energy emission in OJ 287, we used the PL model to describe the emission from the relativistic jet and added the \texttt{optxagnf} model with its PL fraction set to be zero to represent the soft X-ray excess component. The soft X-ray excess is hence deduced to be from an optically thick, low-temperature thermal Comptonisation of the disc emission (section \ref{sec:excess}). Coincidentally, the soft component in our model was consistent with that extracted from our C3PO method, which indicates that it is relatively stable compared to the overall variable PL emission. We did not find an obvious time lag from the soft band in the DCF analysis of the epoch 3 (Figure \ref{fig:dcf}), but this is not surprising because this soft component comprises only ~13$\%$ of the whole emission in 0.5 -- 2 keV and the time bins are relatively large (5752 s) to detect small time lags. 

The time span of epoch 3 is long, $\sim$750 ks, and presents visible variations in both soft and hard bands (Figure \ref{fig:curve}). We divided this epoch into nine segments as delineated in this light curve and extracted their spectra individually. In light of the limited data statistics associated with those much smaller exposure times, we fixed the models to be \texttt{optxagnf+PL}, and kept the parameters of \texttt{optxagnf} to be the same as in Table \ref{tab:optxagn} except for the normalizations, and give those results in Table \ref{tab:norm}. The fits are all acceptable, as shown by the  $\chi^{2}$/d.o.f. values. For each of the nine segments, the normalizations of the \texttt{optxagnf} component are hard to confine but are all consistent within error bars with a median value of $\sim$ 0.45. The variation of the light curve was dominated by the changes in the PL component, with its amplitude as the primary reason (see the dimmed Segments 02 and 07), and its photon index as the secondary one. 

\section{Conclusions}
We explored five pointed long-exposure observation of the blazar OJ 287 made with {\it Suzaku} at three distinct epochs. We searched for temporal and spectral variability on IDV timescales, and for any lags between soft and hard X-ray energies. We studied variations in the HR and performed PSD analyses to see if there was any evidence for the presence of QPOs. We also investigated the origin of soft excess and verified the presence of the soft component from disc.
We can summarize our conclusions as: 

\noindent
\begin{enumerate}
\item[{$\bullet$}] The source showed significant IDV during all three epochs of observations. The fractional variability amplitude ranges from 3.6\% to 11.5\%, and IDV timescales between 16.6 hours to 30.8 hours in the total X-ray energy range of 0.5 -- 10 keV were estimated. \\
\item[{$\bullet$}] The DCF analysis between soft (0.5 -- 2 keV) and hard (2.0 -- 10 keV) showed they are well correlated, peaking at zero lag during epochs 2 and 3 of these observations. These DCFs indicate that the X-ray emission in the soft and hard X-ray bands are predominantly cospatial and emitted from the same population of leptons. \\
\item[{$\bullet$}] We did not find any significant variation in the HR during any epoch of observation. \\
\item[{$\bullet$}] We performed PSD analyses on the total energy LCs during all three epochs of observations of OJ 287. We found these PSDs to be red noise dominated, and no significant QPO was detected during any of these observations. \\
\item[{$\bullet$}] Spectral analysis of all three epochs of observations were carried out and each spectrum was fit with PL models. We found that during epochs 1 and 2, the PL model was the best fit, whereas PL model was not good during epoch 3 because of the soft excess. \\
\item[{$\bullet$}] Using the model independent method C3PO, we found a soft X-ray excess in the spectra of Epoch 3 with no big time lag between hard and soft
bands. While with most of the contribution is due to the PL component, the soft X-ray excess contribution is $\sim 13\%$ below 2 keV and is  well accounted for by the \texttt{optxagnf} model.\\
\item[{$\bullet$}] The almost flat X-ray spectrum of the long 2015 observation indicates a thermal component from the accrection disc that contributes to the emission in the soft X-ray band, while the emission in the whole band is dominated by synchrotron emission from the jet. The thermal component from the accretion disc was seen in this low-activity post-impact state of the 2013 impact but was not seen in either the low-activity pre-impact state nor in the high activity post-impact state in 2007.
Using synchrotron losses and observed timescales, we estimated an upper bound for the size of the emitting region and a lower bound for the typical magnetic field and found that these constraints are consistent with those inferred from broadband SED studies.

\end{enumerate}

\section*{ACKNOWLEDGMENTS}
We thank the anonymous reviewer for several useful suggestions. This research has made use of data obtained from the {\it Suzaku} satellite, a collaborative mission between the space agencies of Japan (JAXA) and the USA (NASA). \\
\\
This work is funded by the National Science Foundation of China (grant no 12233005). ACG is partially supported by Chinese Academy of Sciences (CAS) President's International Fellowship Initiative (PIFI) (grant no. 2016VMB073). PK acknowledges support from the Department of Science and Technology (DST), Government of India, through the DST-INSPIRE faculty grant (DST/INSPIRE/04/2020/002586). MFG is supported by the National Science Foundation of China (grant 11873073), Shanghai Pilot Program for Basic Research–Chinese Academy of Science, Shanghai Branch (JCYJ-SHFY-2021-013), the National SKA Program of China (Grant No. 2022SKA0120102), and the science research grants from the China Manned Space Project with NO. CMSCSST-2021-A06. HGX is supported by the Ministry of Science and Technology of China (grant nos 2020SKA0110201, 2020SKA0110102) and the National Natural Science Foundation of China (grant no 12233005). \\
\\
\section*{Data Availability}
The data used in our paper are publicly available in the Data ARchive and Transmission System for \Suzaku: [https://darts.isas.jaxa.jp/astro/suzaku/data/]. The data underlying this article will be shared on reasonable request to the corresponding author. The software used for data analysis is HEADAS \citep[v6.29c;][]{1995ASPC...77..367B}.\\
\\



\bibliographystyle{mnras}
\bibliography{paper} 





\bsp	
\label{lastpage}
\end{document}